\documentclass[twocolumn,showpacs,preprintnumbers,amsmath,amssymb]{revtex4}
\usepackage{graphicx}% Include figure files
\usepackage{dcolumn}% Align table columns on decimal point
\usepackage{bm}% bold math

\newcommand{\ket}[1]{\left| #1 \right>} % for Dirac kets
\begin{document}

\preprint{in preparation for Phys. Rev. B}

\title{Hyperfine interaction mediated exciton spin relaxation in (In,Ga)As quantum dots}
\author{H. Kurtze$^1$, D.R. Yakovlev$^1$, D. Reuter$^2$, A.D. Wieck$^2$, and M. Bayer$^1$}
\affiliation{$^1$ Experimentelle Physik 2,
                Technische Universit\"at Dortmund,
                D-44221 Dortmund, Germany}
\affiliation{$^2$ Angewandte Festk\"orperphysik,
             Ruhr-Universit\"at Bochum,
             D-44780 Bochum,
             Germany}
\date{\today}

\begin{abstract}
The population dynamics of dark and bright excitons in (In,Ga)As/GaAs quantum dots is studied by two-color pump-probe spectroscopy in an external magnetic field. With the field applied in Faraday geometry and at $T$$<$20~K, the dark excitons decay on a ten nanoseconds time scale unless the magnetic field induces a resonance with a bright exciton state. At these crossings their effective lifetime is drastically shortened due to spin flips of either electron or hole by which the dark excitons are converted into bright ones. Due to the quasi-elastic character we attribute the origin of these flips to the hyperfine interaction with the lattice nuclei. We compare the exciton spin relaxation times in the two resonances and find that the spin flip involving an electron is approximately 25 times faster than the one of the hole. A temperature increase leads to a considerable, non-monotonic decrease of the dark exciton lifetime. Here phonon-mediated spin flips due to the spin-orbit interaction gradually become more important.\end{abstract}

%Without external field, a temperature increase leads to a considerable, non-monotonic decrease of the dark exciton lifetime due to enhanced phonon-mediated spin flips so that above 100~K no corresponding population is found. An enhancement in this reduction at about 60~K indicates that at this  temperature a spin-flip event becomes possible which is most likely related to hole excitation into an excited state.

% , the Physics and Astronomy
%71.70.-d    Level splitting and interactions
%...71.70.Ej    Spin–orbit coupling, Zeeman and Stark splitting, Jahn–Teller effect
%...71.70.Gm    Exchange interactions
%%%%%%%%%%%%%%%%%%%%%%%%%%%%%%%%%%%%%%%%%%%%%%%%%%%%%%%%%%%%%%%%%%%%%%%%%%%%%%%%
%78.20.-e    Optical properties of bulk materials and thin films
%...78.20.Ls    Magnetooptical effects
%%%%%%%%%%%%%%%%%%%%%%%%%%%%%%%%%%%%%%%%%%%%%%%%%%%%%%%%%%%%%%%%%%%%%%%%%%%%%%%%
%78.55.-m    Photoluminescence, properties and materials
%...78.55.Cr    III–V semiconductors
%%%%%%%%%%%%%%%%%%%%%%%%%%%%%%%%%%%%%%%%%%%%%%%%%%%%%%%%%%%%%%%%%%%%%%%%%%%%%%%%
%78.67.-n    Optical properties of low-dimensional, mesoscopic, and nanoscale materials and structures
%...78.67.Hc    Quantum dots
%%%%%%%%%%%%%%%%%%%%%%%%%%%%%%%%%%%%%%%%%%%%%%%%%%%%%%%%%%%%%%%%%%%%%%%%%%%%%%%%


\maketitle

\section{Introduction}
\label{SecI}

The spin dynamics in quantum dots (QDs) have attracted considerable interest recently as they strongly differ from the dynamics observed in systems of higher dimensionality.\cite{Dyakonov} For QD-confined carriers the spin-orbit (SO) interaction and the hyperfine (HF) interaction have been identified as relevant factors for these dynamics.\cite{Khaetskii2001,Merkulov2002,Khaetskii2002,Coish2004,Golovach2004,Petta2005} The relaxation processes can be characterized by the longitudinal relaxation time~(T$_1$) and the transversal relaxation time~(T$_2$).\cite{Slichter} T$_1$ describes the relaxation between the spin states split by an external magnetic field, while T$_2$ describes the damping of the precessional motion about the magnetic field. Here we focus on T$_1$-relevant scattering processes.

Relaxation processes induced by the two prime mechanisms for spin scattering carry different characteristics: Due to the small nuclear Zeeman splitting in the sub-$\mu$eV range flip-flop processes with carrier spins mediated by the HF interaction are quasi-elastic so that they require quasi-degeneracy of the involved carrier spin levels. By contrast SO-mediated scattering events involve acoustic phonons which have substantial density of states and significant interaction matrix elements at larger splittings in the order of a meV. %of interaction with QD-confined carriers

The consequences of these characteristics have been studied so far mainly for single QD-confined carriers, electrons or holes. At $B$=0 the SO interaction is inefficient for carriers in the degenerate ground state Kramers doublet, leading to spin relaxation times up to milliseconds at cryogenic temperatures.
\cite{Khaetskii2001,Woods2002,FinleyElektron,FinleyLoch}
% Fras zitieren -- fertig.
When applying a sufficiently strong magnetic field, though, phonon induced relaxation can be strongly enhanced, leading to a dramatic shortening of T$_1$. Otherwise, as long as the field-induced Zeeman energy is comparable to the nuclear spin splitting, the HF interaction is the dominating relaxation mechanism.\cite{Braun2005,Fras2011}

The situation is distinctly different for charge neutral excitons, which are the focus of interest here. For them the electron-hole exchange interaction leads to a finite energy splitting between bright and dark excitons (details below), so that phonon mediated relaxation may occur already at zero magnetic field.\cite{Gammon2001,Paillard2001,Tsitsishvili2002}

The importance of the HF interaction for exciton complexes is not so clear yet. For bright excitons, which couple to the light field, the lifetime of about a nenosecond is shorter than the time during which this interaction can act efficiently. Dark excitons have considerably longer lifetime up to microseconds, and hence were investigated recently with respect to potential applications in quantum information.\cite{Skolnick10,Fallahi10} Especially for them the HF interaction might become relevant.

Generally, the HF interaction can be decomposed into three contributions: (1) the Fermi-contact interaction, (2) the magnetic dipole-dipole interaction, and (3) the orbital momentum coupling to the nuclei.\cite{Coish2008} These interactions have to be evaluated for the wave functions of an electron in the conduction band or a hole in the valence band. The conduction band in zinc-blende semiconductors is formed by s-type orbitals, so that only interaction (1) is important for electrons, while the other two terms vanish when evaluating the corresponding matrix elements.

In contrast to these findings for the conduction band, the HF interaction of holes has been thought to be much weaker, if not
negligible: The p-type Bloch function results in a vanishing carrier
density at the nuclear sites. If so, the holes might have relaxation
times exceeding those of the
electrons.\cite{OpticalOrientation,BulaevLoss0507,Laurent2005}
However, recent theoretical works have gone beyond the
simplification down to the Fermi-contact terms and have shown that
the anisotropic parts of the HF interaction may contribute
significantly to the hole spin dynamics. The hyperfine interaction
strength between a hole spin and the nuclear spins may be comparable
to the electron hyperfine interaction, depending on the valence band
structure of the QD ground state which is determined by strain and
confinement.\cite{Coish2008,Koudinov2004} If the valence band ground state contains light-hole admixtures, the interaction Hamiltonian is not of Ising-type any longer and hole spin flips become possible. Very recent studies have validated these estimations and have shown that the ratio of the hole and electron HF interaction strength is in the order of 0.1.\cite{Eble2009,Testelin2009,Skolnick10,Fallahi10}

In this paper we study the spin relaxation in self-assembled In(Ga)As/GaAs QDs. By means of magnetic fields we tune the exciton fine structure into resonances where quasi-elastic spin flips can take place between dark and bright ground-state excitons. For the low-temperature regime, the spin flips are initiated by the HF interaction where the nuclei serve as scattering partners. The exciton conversions require a spin flip of either an electron or a hole. Our results show that we can separate these two flip-flop processes so that we can distinguish quantitatively between the electron HF interaction strength and that of a hole. Spin relaxation processes due to the SO interaction are also examined.

\begin{figure}[t]
\includegraphics[width=\linewidth]{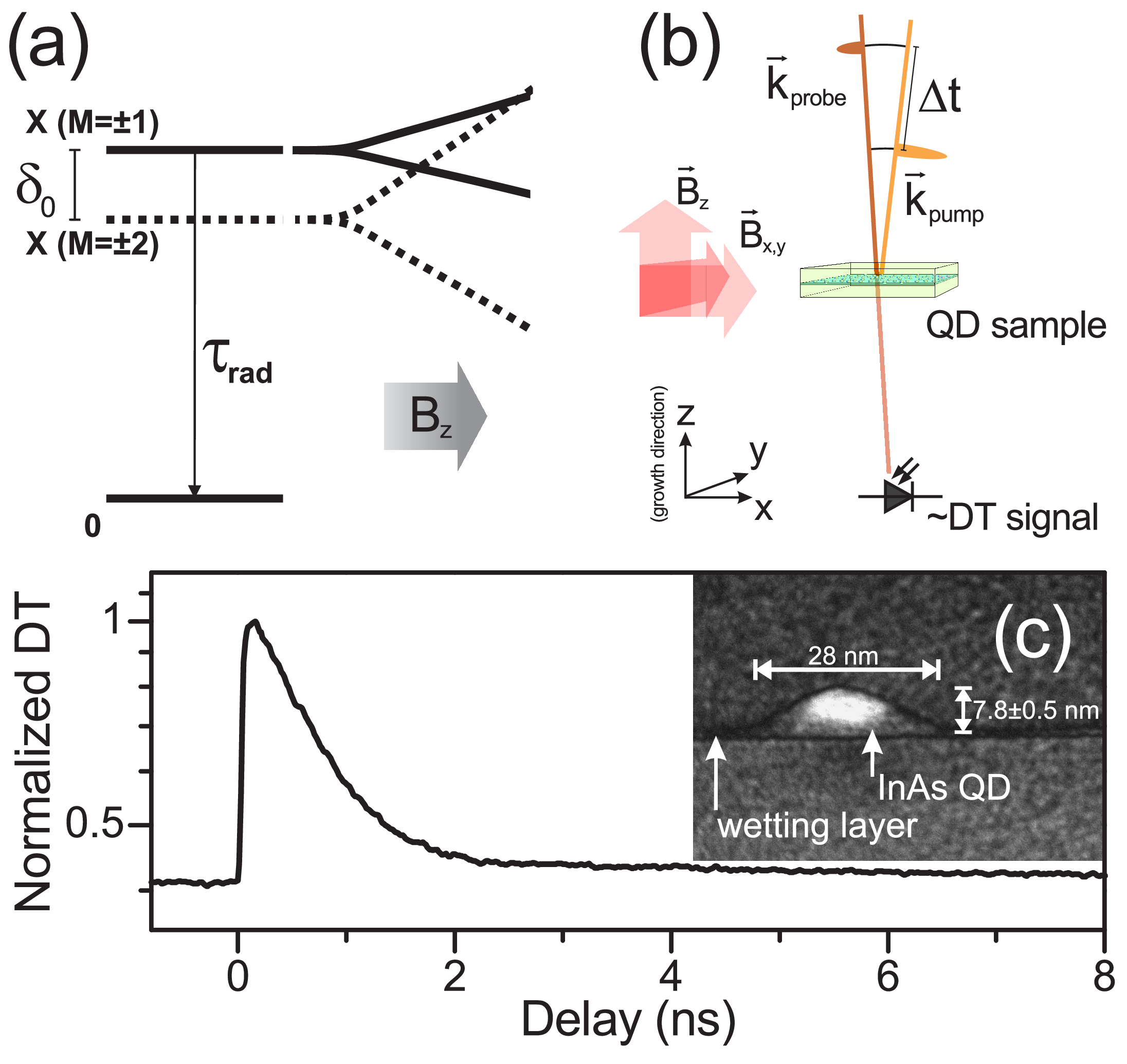}
\caption{(Color online) (a) Level diagram of bright (angular
momentum $M$$=$$\pm1$) and dark ($M$$=$$\pm2$) excitons for the QDs under
study, subject to an external magnetic field B$_z$ along the QD growth
direction. $\delta_0$ denotes the (isotropic) exchange interaction
energy; $\tau_{\textrm{rad}}$ is the radiative decay time. (b) Sketch of the experimental
configuration as described in the text. (c) Time-resolved DT trace,
observed for a pump (probe) excitation density of
$I_0$$=$10~W/cm$^2$ (1~W/cm$^2$) at $T$$=$10~K. Note the log$_{10}$ scale for the vertical axis. The inset gives a cross-sectional
transmission electron micrograph image of an unannealed
self-assembled InGaAs/GaAs QD, nominally identical to the as-grown
QDs in this work. }\label{fig:A1}
\end{figure}

%%%%%%%%%%%%%%%%%%%%%%%%%%%%%%%%%%%%%%%%%%%%%%%%%%%%%%%%%%%%%%%%%%%%%%%%%%%

\section{Experimental technique and sample characterization}
\label{SecII}

The spin dynamics addressed here is distinctly different from that in QDs charged only by a single carrier species because of the energy level structure: A ground state exciton is formed by an electron with spin $S_{e,z}=\pm1/2$ and a hole with $J_{h,z}=\pm3/2$, assuming pure heavy hole character. The exchange interaction couples the electron and the hole spin, resulting in total angular momentum projections $M=S_{e,z}+J_{h,z}=\pm1$ and $\pm2$ which correspond to bright and dark excitons, respectively. These states are split by the isotropic exchange $\delta_0$.

For the dots under study, the $\delta_0$ splitting amounts to $\approx$100~$\mu$eV which would have to be released for spin flips between the two exciton reservoirs. Applying a magnetic field induces a Zeeman splitting of the bright and the dark excitons so that resonances between dark and bright exciton states can occur at particular longitudinal fields [as seen also in the scheme in Fig.~\ref{fig:A1}~(a)].

The experiments are performed by a two-color pump-probe technique
using two synchronized, independently wavelength-tunable Ti:sapphire
lasers. The lasers emit linearly polarized pulses of 1.5~ps duration
at a repetition rate of 75.6~MHz. The temporal jitter between the
two pulse trains is well below 1~ps. We use one laser as a pump which
excites carriers non-resonantly in the GaAs barrier. The
other probe laser is used to test the exciton population in the QD ground
state. The non-resonant pump excitation ensures that spin
relaxation occurs during relaxation of
carriers into their ground states (characterized by a time scale on
the order of 10 ps). Therefore quantum dots capturing an
electron-hole pair can contain exciton spin configurations that are
bright or dark.

The temporal delay~$\Delta$$t$ between pump and probe is adjusted by
a micrometer-precise mechanical delay line. The resulting
time-resolved differential transmission (DT) signal is detected by
a pair of balanced Si-photodiodes connected to a lock-in amplifier,
by which we take the difference between the probe beam sent through
the sample with and without pump action. The pump excitation density
into the GaAs barrier at 1.55~eV photon energy is
$I_0$$=$10~W/cm$^2$. The probe density is chosen to be ten times
weaker at an energy matching the center of the QD ground state emission band (1.37~eV). At the used pump
densities we detect basically only QD ground state emission in
photoluminescence (PL), limiting the number of electron-hole pairs per dot to a
maximum of two. Comparing emission spectra for different excitation
powers shows further that the average exciton occupation per QD is
well below unity at the applied pump power, as the ground state
emission intensity is less than half of its saturation level.

Figure~\ref{fig:A1}~(b) shows a sketch of the experimental
configuration. We apply magnetic fields $B$$\leq$7~T either in the
longitudinal Faraday configuration (parallel to the sample growth
direction and the optical axis ${\bf z}$) or in the transverse Voigt
configuration (perpendicular to ${\bf z}$). The fields are generated
by an optical split-coil magnetocryostat. The QD sample under study is placed in the variable temperature
insert of the cryostat which allows us to lower the temperature down to
5~K.

The heterostucture was grown by molecular beam epitaxy and
contains 10 layers of nominally undoped (In,Ga)As/GaAs QDs,
separated from each other by 100-nm-wide barriers. To get an idea of
the dot geometry, a QD sample was studied by high resolution
transmission electron microscopy [see inset in
Fig.~\ref{fig:A1}~(c)]. From this micrograph we estimate the dot
dimensions to be about 8~nm in height and 30~nm in diameter. The
structure was exposed to post-growth rapid thermal annealing (RTA, 30 seconds
at 920$^{\circ}$C), leading to interdiffusion of dot and barrier material and enhancing the QD volume. As a result the QD emission is
shifted into the sensitivity range of the used Si detectors.

Figure~\ref{fig:A1}~(c) shows a typical DT trace at $T$$=$10~K. The
non-resonant pump excitation at $\Delta t = 0$ excites carriers
which quickly relax to the dot ground state leading to a fast
rise of the DT signal on a 10~ps time scale. The subsequent time
evolution shows two components decaying on different time scales.
The first component shows a fast drop with 0.4~ns time constant. We attribute it to bright exciton decay, as the
same time is observed for the emission decay $\tau_{\textrm{rad}}$ in time-resolved PL
(not shown). The slow component decays on times of about 6$\pm$1~ns,
so that a fraction of this population is still present when the next
pump pulse hits the sample (DT$>$0 for negative delays). Therefore
we associate this population with dark excitons formed by spin flips while relaxing toward the QD ground state after excitation.

\section{Experimental Results}
\label{SecIII}

The experimental results are presented and discussed in the following order: First we analyze exciton spin relaxations by the DT evolution under transverse and longitudinal magnetic fields (subsections A and B, respectively). %In the latter case  resonances in the fine structure between dark and bright ground-state excitons mentioned above. For the low temperature regime we find  that the resonances are initiated by the HF interaction. At higher temperatures one- and especially two-phonon assisted spin flips induced by the SO interaction become relevant (C).
Next, we weigh up the HF and the SO interaction at different temperature regimes (C). Finally the results on the spin relaxation due to the HF interaction are examined and discussed (D).

\subsection{Exciton spin relaxation in transverse magnetic fields}
\label{SecIIIa}

\begin{figure}[t]
%Requires \usepackage{graphicx}
%\centering
\includegraphics[width=\linewidth]{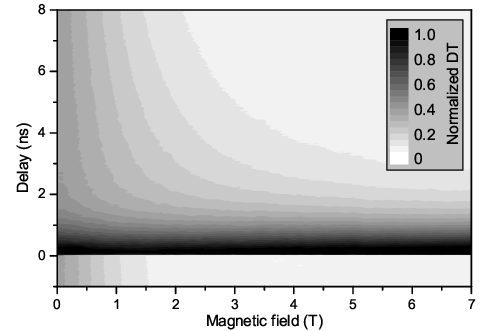}
\caption{Contour plot of DT traces (along the vertical axis) vs
applied magnetic field in Voigt configuration (\textbf{B} $\perp$ \textbf{z}, along the horizontal axis). $T$$=$10~K, $I_0$ pump
power.}\label{fig:voigt}
\end{figure}

The experiments in Voigt configuration confirm our bright and dark exciton assignment made above. In transverse magnetic fields the dark excitons are mixed with the bright excitons, as the rotational symmetry about the growth axis is broken so that the oscillator strength is distributed among the four exciton states.\cite{Bayer2000} This leads to an enhanced decay rate of the dark excitons. The mixing is the stronger the higher the applied magnetic field is and consequently the dark exciton population decreases with increasing $B$. The DT data as shown in Fig.~\ref{fig:voigt} reflect this process. The contour plot consists of DT traces (along the vertical axis) as function of the applied Voigt field strength (horizontal scale) up to 7~T, recorded in steps of 0.2~T. The DT amplitude due to the dark exciton population at larger delays $\Delta$$t$ decreases smoothly with increasing magnetic field, and for field strengths exceeding $\sim$5~T has basically vanished completely.

%%
% HFI
%%

\subsection{Exciton spin relaxation in longitudinal magnetic fields}
\label{SecIIIb}

Let us now focus on studies in longitudinal magnetic fields. Following the notations in Ref.~\onlinecite{Yugova}, the exciton fine structure is given by the effective spin Hamiltonian:
\begin{equation}
{\cal H}_X = \mu_B\left(g_{e,z}S_{e,z}+\frac{g_{h,z}}{3}J_{h,z}\right)B -
\frac{2}{3} \delta_0 S_{e,z} J_{h,z},
\end{equation}
where $\mu_B$ is the Bohr magneton and the $g_{e,z}$ and $g_{h,z}$ are the electron and hole $g$-factors along ${\bf z}$. The anisotropic exchange-splittings of the bright excitons (typically denoted by the energy $\delta_1$) and of the dark excitons ($\delta_2$) are neglected here because $\delta_0\gg\delta_1,\delta_2$, see Ref.~\onlinecite{Bay99}.

\begin{figure}[]
%Requires \usepackage{graphicx}
%\centering
\begin{center}
\includegraphics[width=\linewidth]{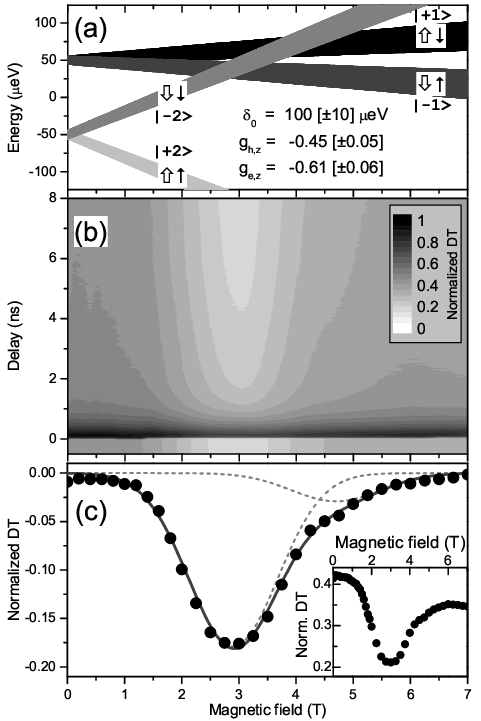}
\caption{(a) Fine structure of bright and dark
exciton states in the studied QDs subject to a longitudinal magnetic
field. The arrows represent the exciton spin configurations where
the electron spins $S_{e,z}= +1/2$ and $-1/2$ are symbolized by thin
arrows pointing up and down, respectively. The thick arrows give the
corresponding orientations of the hole spin $J_{h,z}=\pm 3/2$. (b)
Contour plot of DT traces vs Faraday magnetic field strength.
$T$$=$10~K, pump power $I_0$. (c) Dots: DT values from (b), averaged over a delay time interval of 150~ps before pump arrival
vs magnetic field. The dots in the main panel (inset) show DT values with (without) a baseline subtraction. The solid line is a fit to the DT data by two
Gaussians, both of 1.56~T width for each resonance, with each Gaussian shown by the dashed
lines.}\label{fig:faraday}
\end{center}
\end{figure}

Equivalent QDs to the ones studied in this work were investigated recently by pump-probe Faraday rotation spectroscopy to determine the parameters of the exciton fine structure Hamiltonian with high accuracy. Besides the exchange interaction $\delta_0=100\pm10$~$\mu$eV, also the electron and hole longitudinal $g$-factors of $g_{e,z}=-0.61$ and $g_{h,z}=-0.45$ were measured.\cite{Yugova} We plot the resulting exciton fine structure splitting as a function of the magnetic field~$B_z$ in Fig.~\ref{fig:faraday}, panel~(a).

The longitudinal field configuration does not break the rotational symmetry so that the exciton angular momentum $M$ remains a good quantum number. The $B$-linear splitting of bright and dark excitons leads to two crossings in the magnetic field dispersion: The energy dispersion of the $\ket{{-2}}$ exciton (consisting of a spin-down hole and a spin-down electron, represented by ${\Downarrow}$ and $\downarrow$, respectively) crosses that of the $\ket{{-1}}$-exciton (${\Downarrow}\uparrow$) around 3~T. It also crosses the $\ket{{+1}}$-exciton (${\Uparrow}\downarrow$) at approximately 4~T.

Figure~\ref{fig:faraday}~(b) shows a contour plot of DT transients as a function of magnetic field ($T$$=$10~K and excitation power $I_0$). At low fields $B_z$$<$1~T the signal shows the two-component behavior already discussed in relation with Fig.~\ref{fig:A1}, panel~(c). After pump action the fast bright population decay is followed by the significantly slower dark exciton decay. For increasing fields up to about 3~T, however, the slow decay component shortens which results in decreasing DT amplitudes for longer delays $\Delta$$t$. It increases again reaching times almost as long as at zero field for even higher fields beyond 6~T.

The center of the resonance almost exactly occurs at the field strength at which the crossing of the $\ket{-2}$ exciton with the $\ket{-1}$ exciton was calculated. Hence the field-resonant reduction of the dark exciton population is attributed to a quasi-resonant spin-flip process. Here dark excitons are converted into bright excitons and vice versa leading to reduced long-delay DT values compared to those at $B$$=$0.\cite{Remark:Spin12}
If the $B_z$-dependent fine structure splitting contained a single coincidence, one would expect the resonance to be symmetric with respect to the crossing point. However, in experiment we find a clear asymmetry toward higher~$B$. This asymmetry may indicate that there is another spin-flip resonance at the crossing point of the $\ket{-2}$ exciton with the $\ket{+1}$ exciton.

To analyze this asymmetry in more detail, the inset in Fig.~\ref{fig:faraday}~(c) shows DT values for a fixed time period vs magnetic field. In order to concentrate on the dark excitons, the time window was chosen to be as late as possible at negative delays $\Delta$$t$$<$0 right before the pump pulse action. The duration of the averaged period is 150~ps to achieve a good signal-to-noise ratio. Besides the clearly resolved resonances, a continuous reduction can be seen which is most likely related to $\ket{+2}$ excitons. These excitons decay faster with increasing $B$ due to increasing energy separation and increasing phonon density which demonstrates that the SO interaction is effective also for low temperatures (see below). In the main panel the resulting dataset is baseline-subtracted in order to remove the smooth variation of the dark exciton lifetime with magnetic field and to focus fully on the resonances. Values below zero indicate a field-induced reduction of the dark exciton population. Besides the main resonance at 2.9~T, a shoulder is observed toward higher magnetic fields, supporting that indeed two field resonances occur. We fit the data with a superposition of two Gaussians each with a halfwidth of 1.56~T. The Gaussian form is justified by the inhomogeneous broadening of the fine structure parameters in the QD ensemble. The data can be well described by this fit as shown by the solid line. The two dashed curves give the individual resonances. Within the experimental accuracy, the field position of 4.7~T for the weak resonance is in accord with the crossing point of the $\ket{-2}$ and $\ket{+1}$ excitons in Fig.~\ref{fig:faraday}~(a) (the thickness of the lines reflects the experimental variation of the fine structure parameters).

%%
% SOI
%%
\subsection{Estimation of the relevant spin relaxation mechanism}
\label{SecIIIc}

The influence of SO-induced spin flips on the exciton evolution is limited at low temperatures: In the single-phonon case a phonon would be absorbed to transfer a dark exciton into a bright one. In perturbation theory the corresponding transition rate is given by (a) the magnitude of the matrix element, (b) the phonon density of states, and (c) the phonon occupation.\cite{Tsitsishvili2003,Tsitsishvili2010} In our case the $\ket{-2}$ exciton comes into resonance with the bright excitons. Though the thermal occupation will be sufficient at $T$$=$10~K, the matrix element and the phonon density of states is about zero due to the small energy splitting resulting in inefficient scattering by SO-processes. However, we see an effective lifetime reduction of the dark excitons in the experiment. % which can be better understood by quasi-elastic spin flip processes between exciton populations under HF assistance as explained above.  %where the nuclei serve as scattering partners.% with a nuclear Zeeman splitting in the 100~neV-range..
%Moreover, phonon-assisted transition rates will be most efficient with the phonon wavelength being comparable to the extension of the carrier wave function. This corresponds to phonon energies in the order of meV which exceeds the Zeeman splitting in our accessible field range.
By contrast the $\ket{+2}$ exciton linearly increases its energy separation from the bright excitons corresponding to increasing matrix elements and phonon densities. Here a smooth variation of the SO-related exciton spin relaxation with magnetic field is expected, but a contribution to the observed resonances should be excluded.

In principle also two-phonon scattering (virtual or real via higher orbitals) may become involved. Such scattering processes were shown to be quite efficient, in particular as they do not require nonzero Zeeman splittings.\cite{BulaevLoss0507,Khaetskii2001,Tsitsishvili2002,Golovach2004,SanJose2006,Trif2009,Roszak2009} However, also these processes cannot account for resonances at particular magnetic fields. Moreover, for real transitions the population of phonon modes that bridge excited QD-exciton states is negligible at T=10~K.

The relative importance of the HF and the SO interaction, however, can be altered by increasing the temperature. Especially real two-phonon scattering processes are expected to contribute at thermal energies significantly higher than $T$$=$10~K. Here the SO interaction might dominate the relevant exciton spin relaxations and as a consequence the HF-related resonances could vanish in experiment.

\begin{figure}[]
%Requires \usepackage{graphicx}
%\centering
\begin{center}
\includegraphics[width=\linewidth]{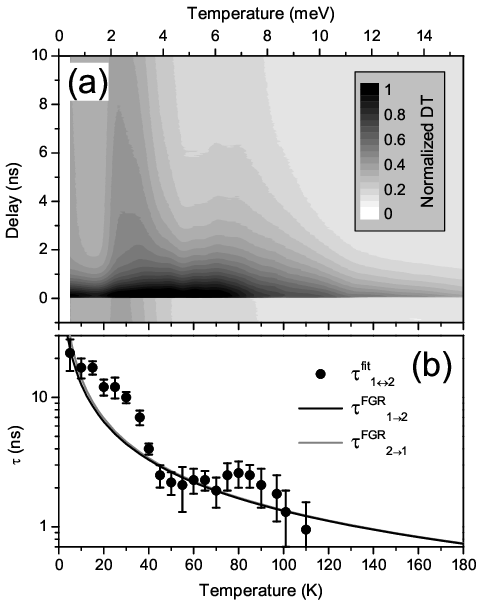}
\caption{(a) Contour plot of normalized DT traces for varying temperatures from 5~K to
180~K; $B$=0, excitation power $I_0$. (b) The data points give relaxation
times~$\tau^{\textrm{fit}}_{1\leftrightarrow 2}$ obtained from a fit to the DT
dataset from panel~(a). Solid lines show calculated single-phonon relaxation times from $M$=${\pm1}$ to $M$=${\pm2}$ excitons (and vice versa) as described in the text (log$_{10}$ scale).}\label{fig:temp}
\end{center}
\end{figure}

Experimental findings confirm these expectations. Figure~\ref{fig:temp}~(a) shows a contour plot made up of DT transients at temperatures between $T$$=$5~K and $T$$=$180~K.  For low temperatures $T$$<$20~K the DT transients reveal the two-component character that was discussed above in relation with Fig.~\ref{fig:A1}~(c). In general with increasing temperatures the difference between the two temporal components tends to fade out and the DT amplitudes for longer delays decrease. However there is a discontinuous behavior in between around 50~K where DT amplitudes for longer delays are less pronounced than at the transients at $T$$=$30 and 70~K.

We analyze the experimental DT data with a rate equation that models the exciton population. Since the transition rates from the energetically higher lying bright $M$$=$$\pm$1 excitons to the lower lying dark $M$$=$$\pm$2 will not differ much from the reverse process at these temperatures, we assume a single spin relaxation time $\tau_{1\leftrightarrow2}^{\textrm{fit}}$ which serves as a fit parameter. (Details are shown in the Appendix.) The resulting evolution was fitted to each DT transient from Fig.~\ref{fig:temp}~(a), and the corresponding relaxation times $\tau_{1\leftrightarrow2}^{\textrm{fit}}$ are given by the dots in panel~(b). The spin-flip times reduce strongly with increasing temperature starting with $\approx$20~ns for $T$$=$5~K down to $\approx$1~ns for 110~K. Beyond that the exciton conversion time shows a plateau between $T$$=$40~K and 80~K where $\tau_{1\leftrightarrow2}^{\textrm{fit}}$ takes on values of approximately 2~ns.

If the underlying spin relaxation mechanism was solely SO-mediated single-phonon scattering, the flip times could simply be described by Fermi's golden rule (FGR).\cite{Tsitsishvili2003,Tsitsishvili2010} For comparison, we compute these flip times $\tau^{\textrm{FGR}}_{1\rightarrow2}$ and $\tau^{\textrm{FGR}}_{2\rightarrow1}$ and plot them as solid lines in panel~(b) of Fig.~\ref{fig:temp}. (Details again in the Appendix.) These one-phonon relaxation times follow the temperature-dependent phonon occupation~$N$ and roughly describe the experimental data. However the results also reveal deviations which indicate spin flips assisted by resonant two-phonon scattering. Here the spin flips involve excitations to higher QD-orbitals so that spin relaxation is possible beyond a distinct thermal activation energy while it is blocked below. Apparently, exciton spin scattering processes are additionally enabled around thermal energies of 4~meV (equivalent to $T$$\approx$40~K; see upper scale of Fig.~\ref{fig:temp}) which corresponds to quantization energies of holes in the valence band.\cite{Remark:Quantization}

\begin{figure}[]
%Requires \usepackage{graphicx}
%\centering
\begin{center}
\includegraphics[width=\linewidth]{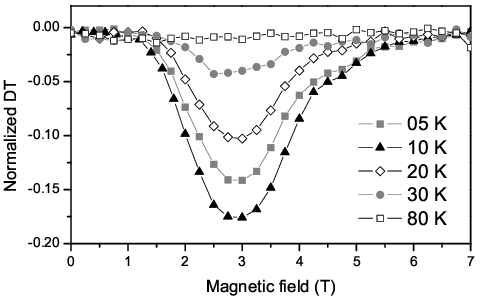}
\caption{DT averaged over a negative delay time interval of 150~ps before pump arrival
vs magnetic field for
various temperatures as indicated. The data are baseline subtracted as in Fig.~\ref{fig:faraday}~(c).}\label{fig:tempreso}
\end{center}
\end{figure}

In fact the SO interaction is the primary exciton spin relaxation mechanism at higher temperatures: Fig.~\ref{fig:tempreso} gives similar data to those of Fig.~\ref{fig:faraday}~(c) but for other temperatures. The field-induced resonances show weak variations at low cryogenic temperatures ($T$$=$5~K to 20~K; small variations in the amplitude could occur due to changes in the generation of initial dark exciton populations). %Moreover, the relative amplitude of the resonance at $B$$=$4.7~T remains comparable to that at $B$$=$3~T, indicating that the underlying HF-relaxation mechanism is not sensitive to temperature changes in the low temperature regime.
But already for $T$$=$30~K the curve shows only relics of the HF-induced resonances. For even higher temperatures the dark excitons are strongly reduced so that the DT curve lacks any magnetic field dependent resonances (DT curve for 80~K).

\subsection{Examination of HF-mediated exciton spin relaxation times}
\label{SecIIId}

\begin{figure}[t]
%Requires \usepackage{graphicx}
%\centering
\begin{center}
\includegraphics[width=\linewidth]{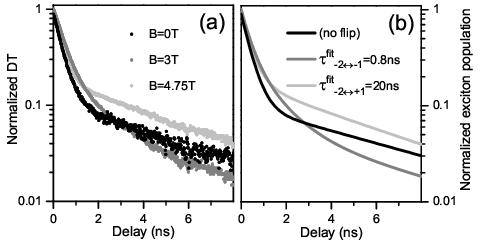}
\caption{(a) Normalized DT values vs delay time at field amplitudes of $B$$=$0, 3~T, and 4.7~T in Faraday configuration (scattered dots). The traces were set to zero at $\Delta$$t$$=$0. (b) Total exciton population as modeled by a set of rate equations; see text. The numbers give the different values of the parameter $\tau^{\textrm{fit}}$ mapping the spin flip of the involved $M$$=$$-$2 dark excitons into $\pm$1 bright ones and vice versa. (Both panels log$_{10}$ scale.)}\label{fig:traces}
\end{center}
\end{figure}

Let us now evaluate the HF-mediated spin-flip times that become relevant when driving the system into the fine-structure determined resonances at low temperatures, $T$$=$10~K. We consider spin conversions between the bright~$M$$=$$\pm$1 and the dark~$M$$=$$-$2 excitons. They are taken into account by spin-flip times $\tau^{\textrm{fit}}_{{-2}\leftrightarrow{\pm1}}$ which serve as a fit parameter to the experimental DT evolution in the set of coupled rate equations mentioned in the section above [other parameters remain unchanged compared to those in relation with Fig.~\ref{fig:temp}~(b)]. Figure~\ref{fig:traces} compares the experimental DT traces [panel (a)] with the exciton evolution modeled by the rate equations [panel (b)].

At $B$$=$0 the exciton populations decay as discussed already above for the undisturbed case (panel~(a)). In comparison the initial decay slows down whereas the decay at longer times becomes faster in the resonances. This behavior is reproduced by the modeled exciton populations (panel~(b)): Right after pump action nonzero spin flip rates convert a fraction of the bright excitons into $\ket{-2}$ excitons by which the fast decay is delayed. For longer delays $\Delta t$, $\ket{-2}$ excitons may undergo spin flips into the bright spin configurations. This is followed by a radiative decay which leads to a reduction of the exciton population compared to the off-resonant case. These changes are most obvious in the case of the electron spin flip at $B$$=$3~T where $\tau^{\textrm{fit}}_{{-2}\leftrightarrow{-1}}$$\approx$(0.8$\pm$0.15)~ns is obtained from the fit.  In contrast, the spin flip time due to a hole flip at B$=$4.7~T is found to be approximately 25~times slower with $\tau^{\textrm{fit}}_{{-2}\leftrightarrow{+1}}$$\approx$(20$\pm$4)~ns.

In the remaining part we examine the HF-mediated spin flip scattering events. In the dominant resonance at $B$$=$3~T excitons are converted into each other where the electron spin ${\bf S}_e$ interacts with one of the nuclei ${\bf I}_i$. This corresponds to an effective Hamiltonian given by
\begin{eqnarray}
{\cal H}_e = \sum_i A_i \left| \Psi \left( {\bf r}_{ei} \right)
\right|^2 {\bf S}_e {\bf I}_i, \label{EQ:electron1}
\end{eqnarray}
where the sum goes over all nuclei in the QD electron localization volume. The interaction strength of the electron spin (${\bf S}_e$) with a nucleus (${\bf I}_i$) is determined by the hyperfine constant $A_i$ specific for each nuclear species in the dot and the electron density $\left| \Psi \left( {\bf r}_{ei} \right) \right|^2$ at the nuclear site ${\bf r}_{ei}$.

The second weaker resonance can be initiated by a hole spin flip only. %(the ${\Downarrow}\downarrow$-exciton crosses with ${\Downarrow}\uparrow$). %Again this flip-process is initiated by the nuclei only, demonstrating the importance of the hole-nuclei HFI.
A dipole-dipole interaction could not convert pure $\pm3/2$ heavy-hole states into each other due to the mismatch of angular momentum exchange.
If it was a pure heavy hole state, the interaction would be described by an Ising-form,
\begin{equation}
{\cal H}_h = v_0 \sum_i C_i \left| \Phi \left( {\bf r}_{hi} \right)
\right|^2 J_{h,z} I_{i,z},\label{EQ:holeIsing}
\end{equation}
where the $J_{h,z}$ are the hole spin projections along the growth axis, the $C_i$ are the interaction constants, and $\left|\Phi \left( {\bf
r}_{hi} \right) \right|^2$ is the hole density at a particular nuclear site.

But in the studied QDs the in-plane hole $g$-factor differs considerably from zero, $g_{h,\perp} = 0.15$.\cite{Yugova} This indicates that the hole ground state contains admixtures of light-hole states $\Phi_{\pm1/2}$ with $J_{h,z}=\pm1/2$ and the mixed hole states are $\tilde{\Phi}_{\pm3/2}  =  \left(\Phi_{\pm3/2} + \beta\Phi_{\mp1/2}\right)/\sqrt{1+\left|\beta\right|^2}$ with the complex mixing coefficient $\beta$. Now hole-nuclei flip-flops become possible and the two resonances can be treated analogously. The interaction Hamiltonians of the two carrier types with a single nucleus~$i$ can be rewritten as
\begin{eqnarray} {\cal H}_{e,i}&=&\frac{v_0}{2} A_i \left| \Psi \left( {\bf r}_{ei} \right) \right|^2\nonumber\\
&\times&\left( \frac{1}{2}(S_{e,+}I_{i,-})+\frac{1}{2}(S_{e,-}I_{i,+})+S_{e,z}I_{i,z}
\right)\label{EQ:electronLadder}
\end{eqnarray}
and
\begin{eqnarray} {\cal H}_{h,i}&=&\frac{v_0}{2} C_i \left| \tilde{\Phi} \left( {\bf r}_{hi} \right) \right|^2\nonumber\\
&\times&\left(
\frac{1}{2}(J_{h,+}I_{i,-})+\frac{1}{2}(J_{h,-}I_{i,+})+S_{e,z}I_{i,z}
\right)\label{EQ:holeLadder}
\end{eqnarray}
with the spin ladder operators $S_{e,\pm}$, $J_{h,\pm}$, and $I_{i,\pm}$. Flip-flop processes are initiated by the first two terms on the right-hand side.

An estimate of the resulting spin flip rates can be obtained from lowest-order perturbation theory. Scattering between the two exciton populations is only possible when the energy difference of the exciton states coincides with the spin splitting of a nucleus, and hence the density of states of the nuclei plays an important role. Unfortunately this parameter is not known with high accuracy but can be at least estimated adequately. At first the nuclear splittings are broadened due to variations of the quadrupole momentum, and second a further broadening arises due to individually deviating magnetic surroundings in the dipole-dipole interaction. For simplicity we assume a uniform density of states that is unspecific with respect to the nuclei and their isotopes. The best agreement with the experimental relaxation times is obtained with all QD-nuclei (N$\sim8.5\times10^4$) spread over 75~neV. This value seems sensible since for (In,Ga)As/GaAs QDs the magnitude of the quadrupole interaction $h\nu_Q$ was estimated to be on the order of 1 to 10~neV, where $\nu_Q$ is the quadrupole frequency resulting from the electric field gradient at a given nuclear site with a certain strain.\cite{Flisinski2010,Cherbunin2011} Also the dipole-dipole interaction will further broaden the energy distribution of the nuclei. Especially central-spin mediated ``co-flips'' between nuclei involving a primary carrier spin might contribute here.\cite{Gammon2001,Cywinski2009} Choosing reasonable values for the remaining parameters describing the QD (details are given in the Appendix),  the calculated spin relaxation times amount to 0.8~ns for an electron flip and 20~ns for a hole flip. Notably the ratio of these two times is in good agreement with our experimental results.

\section{Conclusions}

%In the resonances quasi-elastic spin flip processes between the bright and dark exciton population occur, and

We studied spin relaxations of excitons in InAs/GaAs QDs. In the undisturbed case the exchange interaction separates bright and dark exciton states energetically. Magnetic fields, however, drive the system into resonances where quasi-elastic spin scattering processes can take place. Two of such resonances are observed at each of which the lifetime of the involved dark exciton state is drastically shortened. Here a spin-flip of either electron or hole occurs, converting the dark into a bright exciton. Due to the quasi-resonant character we assign the origin of the underlying scattering process to the HF interaction. As we observe both electron and hole flips their spin dynamics can be compared for a single sample under comparable experimental conditions. The results show that spin flips involving an electron are approximately 25 times faster than those assisted by a hole flip. Our results are in agreement with previous studies on the HF interaction of QD-confined carriers where the HF interaction strengths of holes are found to be one order of magnitude smaller that those of the electrons.\cite{Coish2008,Eble2009,Testelin2009,Skolnick10,Fallahi10}
A comparison with transition rates that are based on first-order perturbation theory verifies our findings. With increasing temperature spin-flips due to the SO interaction gradually become more important in the relaxation processes and finally dominate the exciton dynamics for $T$$>$30~K.

%For the low temperature regime, the resonances are initiated by the HF interaction. We demonstrated that the two resonances can be attributed to an electron- and a hole spin flip with the nuclei. This confirms that the hole spin HF interaction is important, even though the relaxation rate is found to be smaller (by a factor of about 25) than for the electron.

% so that the broadening of the individual nuclei splittings is still well below the nuclear Zeeman splitting at the prevailing fields (e.g. 220~neV for $^{69}$Ga at $B$$=$3.5~T).

\begin{acknowledgments}
The support by the PlusLucis project, Ziel2.NRW program and Deutsche Forschungsgemeinschaft (DFG 1549/10-1) is acknowledged. We thank J.-M. Chauveau, CNRS Valbonne, and A. Ludwig, Ruhr-Universit{\"a}t Bochum (now Department of Physics, University of Basel) for providing the TEM image.
\end{acknowledgments}

\appendix*
\section{Details of the modeled exciton spin relaxation times}
\label{appendixA}

The exciton evolution mentioned in Sec.~\ref{SecIIIc} is modeled by coupled rate equations where M$=$$+1$, $-1$, $+2$, and $-2$ excitons undergo spin flips or decay. The temperature-dependent spin relaxation between the bright and dark exciton population is described by the conversion time $\tau_{1\leftrightarrow2}^{\textrm{fit}}$. Bright excitons decay with the above-mentioned $\tau_{\textrm{rad}}$=0.4~ns obtained by time-resolved PL. The rate equation also assumes a phenomenological non-radiative decay channel for all excitons of $\tau_{\textrm{nonrad}}$$=$8~ns which results together with $\tau_{1\leftrightarrow2}^{\textrm{fit}}$ in the dark exciton decay rate $\approx$6~ns mentioned above for $T$$=$10~K.

Assuming single-phonon scattering, the exciton spin relaxation times can be written according to Fermi's golden rule (FGR) as $$\tau^{\textrm{FGR}}_{1\rightarrow2}  =  \tau_{0} \left(N+1\right)^{-1} \textrm{and}\; \tau^{\textrm{FGR}}_{2\rightarrow1} = \tau_{0} N^{-1},$$ where $N=\left( \exp (\delta_0/kT) \right)^{-1}$ is the phonon occupation factor for a given temperature~$T$ and exchange splitting~$\delta_0$, with $k$ being the Boltzmann constant. All other underlying parameters are summarized in a zero-temperature relaxation time~$\tau_0$.\cite{Tsitsishvili2003,Tsitsishvili2010} Together with the above-mentioned $\delta_0=100$~$\mu$eV, a value of $\tau_0=115$~ns was used.

The transition rate for an initial state $\ket{m}$ into a final state $\ket{n}$ is $\Gamma_{mn}=\frac{2\pi}{\hbar} \varrho(E_n) |\langle {n}|{\cal H}_{e/h}|{m}\rangle|^2$, where $\varrho(E_n)$ is the density of the final states and ${\cal H}_{e/h}$ denotes the Hamiltonians above [Eqs.~(\ref{EQ:electronLadder}) and~(\ref{EQ:holeLadder})]. The parameters for the numerical estimation of exciton spin flip times due to the HI interaction in Sec.~\ref{SecIIId} are as follows: The number of unit cells is obtained via the QD volume. For the annealed QDs not enough contrast is obtained in electron microscopy to get reliable data on their geometry parameters. We assume a QD with a base diameter of 35~nm and a height of 8~nm [cf. the inset of an unannealed QD in Fig.~\ref{fig:A1}~(c)] and hence the modeled QD contains $\sim8.5\times10^4$ nuclei. For reasons of simplicity the carriers are believed to be extended over the whole QD volume with a uniform probability. In order to consider material interdiffusion due to the RTA, a Ga-intermixture of 0.42 is used. Usually the hole is located at the top of the QD, which is indium-rich compared to the bottom, so that in this case a Ga-intermixture of 0.18 was taken into account.\cite{intermixture} The In-, As-, and Ga-specific hyperfine constants as well as the nuclear spin~$I$ are given in the table below. (Values from Ref.~\onlinecite{Testelin2009}.)
\begin{center}
\begin{tabular}{c|c|c|c}
Species & $A_i$($\mu$eV) & $C_i$($\mu$eV) & I \\ \hline
In & 56 & 4.0 & 9/2\\
As & 46 & 4.4 & 3/2\\
Ga & 38 & 3.0 & 3/2\\
%%Lines of table here ending with \\
\end{tabular}
\end{center}
Clearly the spin relaxation rate for a hole depends strongly on the light hole admixture $\beta$. 0.2$<$$\left|\beta\right|$$<$0.7 were reported for strained dots.\cite{valuebeta} For our calculations we use $\left|\beta\right|$=0.5.


\begin{references}
\bibitem{Remark:Spin12} Spin-flips within the bright or dark exciton reservoir are not of interest here since the latter ones are not relevant for the dark exciton lifetime shortening. Additionally such spin flips require relaxation processes of both electron and hole with a strongly reduced probability.
\end{references}
\end{document}